\documentclass[twocolumn,aps,showpacs,prl,amsmath]{revtex4-1}

\usepackage{amssymb}
\usepackage{graphicx}
\usepackage{dsfont}
\usepackage{color,soul}

\begin{document}

\title{Klein tunneling and electron optics in Dirac-Weyl fermion systems \\ with tilted energy dispersion}

\author{V. Hung Nguyen\footnote{E-mail: viet-hung.nguyen@uclouvain.be} and J.-C. Charlier} \address{Institute of Condensed Matter and Nanosciences, Universit\'{e} catholique de Louvain, Chemin des \'{e}toiles 8, B-1348 Louvain-la-Neuve, Belgium}

\begin{abstract}
	The outstanding electronic properties of relativistic-like fermions have been extensively studied in solid state systems with isotropic linear dispersions such as graphene. Here, we show that 2D and 3D Dirac-Weyl (DW) materials exhibiting tilted energy dispersions could induce drastically different transport phenomena, compared to the non-tilted case. Indeed, the Klein tunneling of DW fermions of opposite chiralities is predicted to appear along two separated oblique directions. In addition, valley filtering and beam splitting effects are easily tailored by dopant engineering techniques while the refraction of electron waves is dramatically modified by the tilt, thus paving the way for emerging applications in electron optics and valleytronics. 
\end{abstract}

\maketitle

\textbf{Introduction} - The success isolation of graphene \cite{novo04}, owing to a low-energy linear-dispersion and the relativistic (Dirac) - like fermions, was a breakthrough in the last decade as it offers many promising prospects in both fundamental and applied research \cite{ferr15}. Among many interesting phenomena discovered in graphene, the Klein tunneling \cite{kats06} is a direct evidence and concurrently provides a playground for implementing tests of relativistic quantum dynamics of quasiparticles in a simple experimental situation \cite{pkim17}. In addition, the Dirac fermions in graphene exhibit several behaviors analogous to light rays in optical media \cite{alla11} such as refraction, reflection, and Fabry-P\'{e}rot interferences, making it an ideal platform for electron optics demonstration and novel quantum device development (e.g., see in refs.\cite{kats06,chei07,wilm14,hlee15,chen16,bogg17}).

Soon after the discovery of graphene, the search for other materials hosting relativistic-like particles has become an emerging topic. Actually, the relativistic-like (i.e., Dirac-Weyl) fermions have been explored in several 2D and 3D materials. Silicene, germanene, stanene, graphynes (i.e., 2D carbon allotropes), phosphorene systems, borophene allotropes \cite{bale15,malk12,jkim15,ylu016,zhou14,goer08}, etc., correspond to the 2D form. Examples in the 3D form include Na$_3$Bi, Cd$_3$As$_2$, monopnictide family (NbP, NbAs, TaP, TaAs), HgTe class, LaAlGe, Mo$_x$W$_{1-x}$Te$_2$, SrSi$_2$, Ta$_3$S$_2$, 3D carbon networks \cite{wang12,kliu14,yxu015,hwen15,ruan16,syxu17,ysun15,chan16,huan16,weng15}. Interestingly, the appearance of these Dirac-Weyl (DW) materials resulted in the exploration of several novel phenomena \cite{armi17}, e.g., quantum oscillations \cite{lphe14}, chiral anomaly \cite{zhan16}, chirality-dependent Hall effects \cite{burk14,shen15}, Klein tunneling and electron optics in 3D systems \cite{hill17}, photovoltaic effects \cite{tagu16}, etc.

In contrast to graphene, the DW cones in several materials \cite{jkim15,ylu016,zhou14,goer08,wang12,kliu14,hwen15,ruan16,syxu17,ysun15,chan16,huan16} are anisotropic and could be exceptionally tilted. Remarkably, the tilt character has been shown to affect strongly the electronic transport and the optical properties of these DW fermions \cite{tres15,solu15,zyuz16,yesi17,shar17,stei17,mukh17,qma017,chan17}. For instance, the small (subcritical) tilts give rise to an asymmetric Pauli blockade, resulting in finite free electron Hall response \cite{stei17} and non-zero photocurrents \cite{qma017}. For large (overcritical) tilts as in type II Weyl semimetals \cite{armi17}, the Fermi surface is no longer point-like, inducing other novel phenomena. For example, the Landau-level spectrum is gapped and has no chiral zero mode if the direction of magnetic field is outside the Weyl cone \cite{solu15}. The anomalous Hall conductivity is not universal and can change sign as a function of the tilt magnitude \cite{zyuz16}. It is worth noting that even being an inherent property of materials, the DW dispersion (hence, the tilt character) can be also generated and tuned, e.g., either by strain or nanoengineering techniques \cite{jkim15,ylu016,goer08,ruan16,chan16}.

In this context, questions about the tilting effects on crucial phenomena such as Klein tunneling and electron optics arise. In this Letter, we report on novel transport properties of DW fermions in hetero-doped structures induced by the effects of subcritical tilt. In particular, we find that instead of being observed in a unique direction (i.e., normal incidence) as in the non-tilted case, the Klein tunneling of tilted DW fermions of opposite chiralities is achieved in two separated oblique directions. In addition, interesting phenomena such as anisotropic Fabry-P\'{e}rot resonances, valley filtering and beam splitting effects, and novel electron optics behaviors in DW \textit{p-n} junctions are also predicted. Besides their importance to fully understand the relativistic-like phenomena of DW fermions, our findings could be the basis for novel applications in electron optics and valleytronics. 

\begin{figure}[!t]
	\centering
	\includegraphics[width = 0.46\textwidth]{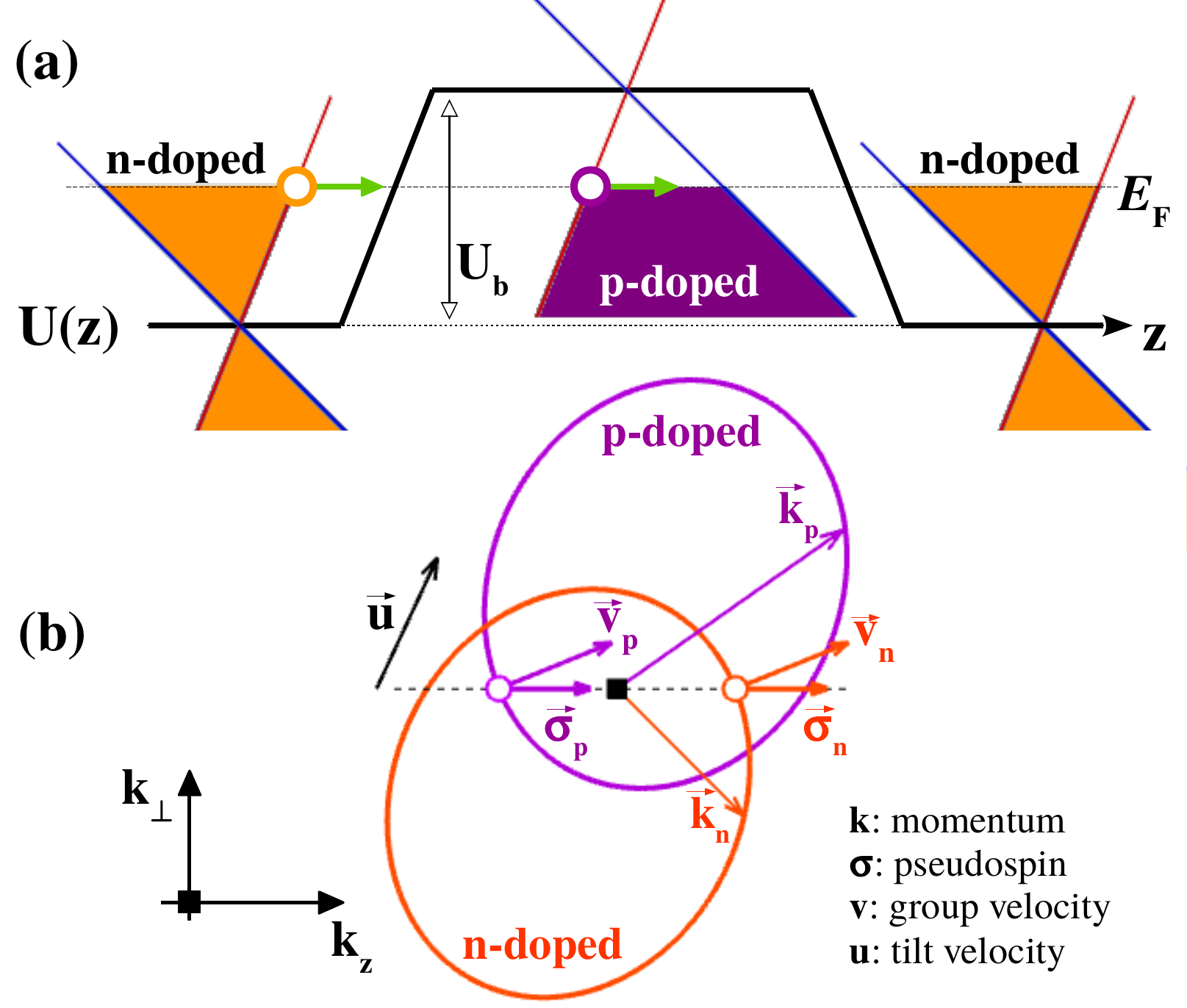}
	\caption{\textbf{Potential barrier in a tilted Dirac-Weyl fermion system}. (\textbf{a}) Energy profile (barrier height: $U_b$) along the transport direction (i.e., Oz axis). (\textbf{b}) Diagram illustrating the momentum shift of Fermi surfaces in the differently doped zones and the transmission process with momentum $\vec k_\perp = 0$. The subscripts \textit{n}/\textit{p} indicate the vectors outside/inside the barrier, respectively, and the black-dashed line implies the $\vec k_\perp$-conservation.}
    \label{fig_sim1}
\end{figure}
\textbf{Methodology} - Around the DW cones, charge carriers can be modeled by a generic Hamiltonian \cite{zyuz16,yesi17,stei17,shar17,mukh17,qma017}:
\begin{equation}
	H_0 = \sum\limits_{\iota = x,y,z} \tau_c  (u_{\iota} p_{\iota} + \sigma_{\iota} v_{F\iota} p_{\iota})
\end{equation}
where $\sigma_{\iota}$ are the Pauli's matrices, $\tau_c$ represents the chirality and $\vec u = (u_x,u_y,u_z)$ describes the tilt of energy dispersion. Since they always appear in pairs with opposite chiralities, $\tau_c = \pm 1$ for the $W$ and $W'$ cones. We consider hetero-doped structures modeled by applying a potential profile $U(z)$ along the Oz axis (see Fig.1) and accordingly the system Hamiltonian reads $H = H_0 + U(z)$. In order to compute the transport properties, we employ the calculation method developed in \cite{hung10}. In particular, $H$ is separated into two independent parts, i.e., $H = H_z + H_\perp$, and then is rewritten in the basis $\{ | z_m \rangle \otimes | \vec k_\perp \rangle \}$ with the mesh-spacing $a_0 = z_{m+1} - z_m$ along the transport direction (i.e., Oz axis) and the plane wave $| \vec k_\perp \rangle $ in the Oxy plane. This Hamiltonian is finally solved using the Green's function technique and the transport properties are extracted (see in Ref. \cite{SM2017} for more details).

The energy eigenvalues of the Hamiltonian (1) are given by $E_{\tau_c\tau_b} (\vec k) = \hbar ( \tau_c \vec u \vec k + \tau_b |\vec \xi_{\vec k}| )$,
where the vector $\vec \xi_{\vec k} = (v_{Fx} k_x, v_{Fy} k_y, v_{Fz} k_z)$ and $\tau_b = \pm 1$ corresponding to the conduction/valence bands, respectively. Accordingly, the pseudospin and group velocity are determined as $ {\vec \sigma}_{\tau_c\tau_b} (\vec k) = \tau_c\tau_b \vec \xi_{\vec k}$ and ${\vec v}_{\tau_c\tau_b} (\vec k) = \tau_c \vec u + \tau_b \vec \eta_{\vec k} / |\vec \xi_{\vec k}|$ with $\vec \eta_{\vec k} = (v_{Fx}^2 k_x, v_{Fy}^2 k_y, v_{Fz}^2 k_z)$ \cite{SM2017}. Some exceptional features, compared to the isotropic dispersion case, are found here. First, the group velocity $\vec v$ and momentum $\vec k$ are no longer collinear. Note that the direction of $\vec v$ (but not of $\vec k$) determines the propagating direction of an electron wave in the crystal. Second, the pseudospin is no longer locked to the direction of electron motion. Third, the center of the Fermi surface in momentum space is energy dependent due to the tilt, offering the possibility to separate and to control this separation of Fermi surfaces in hetero-doped systems as illustrated in Fig.\ref{fig_sim1}.b. These features lay the foundations to observe the novel transport phenomena reported in this Letter. 

Unless otherwise stated, to concentrate on the tilting effects, the velocities $v_{F\iota}$ in Eq.(1) are assumed to be isotropic, i.e., $v_{Fx} \equiv v_{Fy} \equiv v_{Fz} = v_F$. Throughout the work, three vectors $\vec v$, $\vec u$ and $\vec \sigma$ are considered in the same form (i.e., $\vec v = v (\sin\theta\cos\phi,\sin\theta\sin\phi,\cos\theta)$, etc.) and the energies are presented in the unit of $E_0 = \hbar v_F/2a_0$. 

\begin{figure}[!t]
	\centering
	\includegraphics[width = 0.49\textwidth]{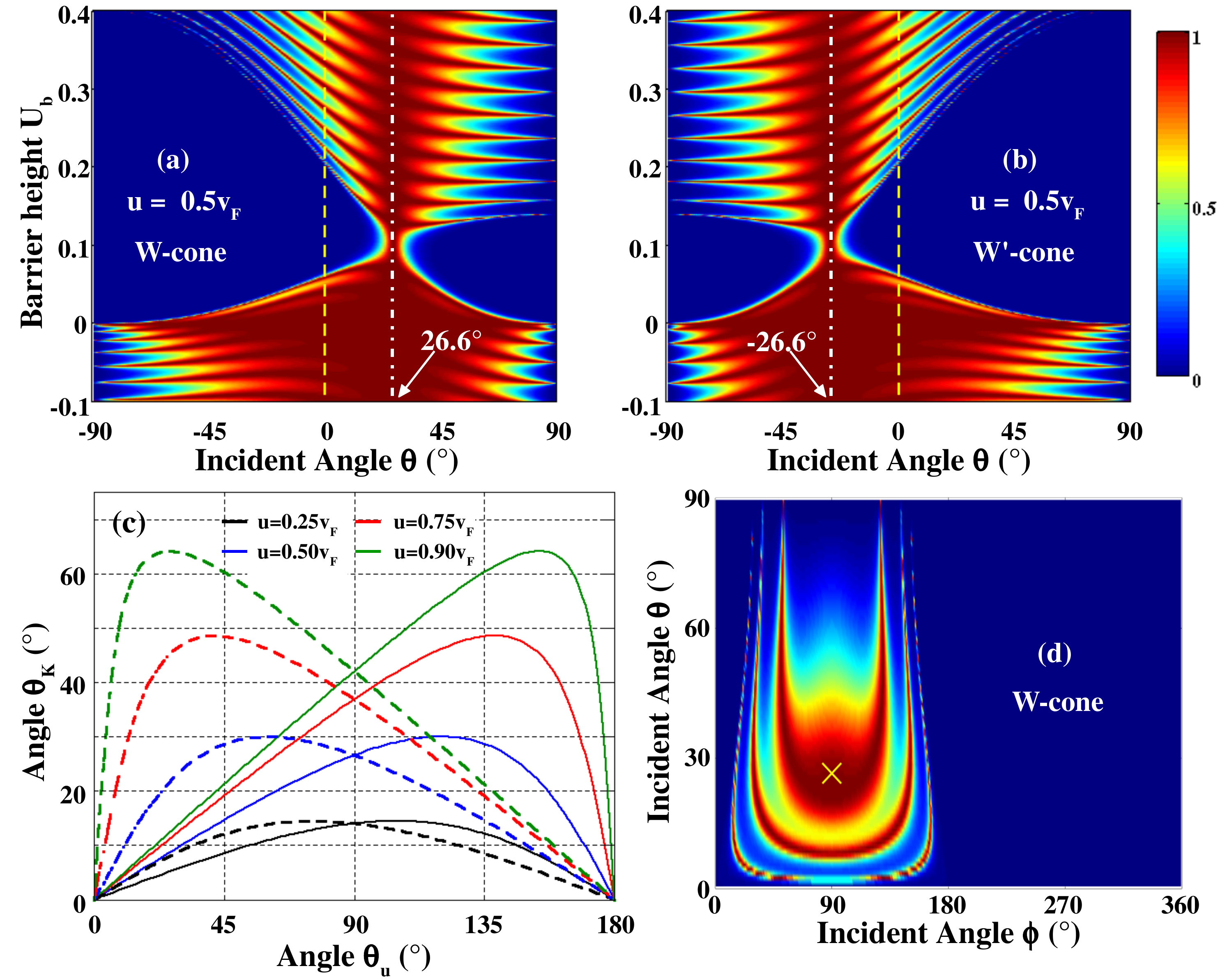}
	\caption{\textbf{Klein tunneling and Fabry-P\'erot resonances.} (\textbf{a,b}) ($U_b, \theta$)-maps of transmission probabilities ($\mathcal{T}_{W,W'}$) of tilted DW fermions in the W and W' cones, respectively. The incident energy $E = 0.1$, the incident angle $\phi = 90^\circ$, the barrier width $L_b = 200 a_0$ while the $\vec u$-direction  $(\theta_u,\phi_u) = (90^\circ,90^\circ)$. (\textbf{c}) Klein tunneling direction $\theta_{K}$ obtained for the $W$ (solid) and $W'$ cones (dashed lines) as a function of $\theta_u$. (\textbf{d}) represents a ($\phi$-$\theta$)-map of $\mathcal{T}_{W}$ at $U_b = 0.2$ in (\textbf{a}). The cross symbol marks the Klein tunneling direction. }
	\label{fig_sim2}
\end{figure}

\textbf{Klein tunneling and Fabry-P\'{e}rot interferences} - Transmission probabilities $\mathcal{T}_{W,W'}$ of tilted DW fermions through a potential barrier (see Fig.\ref{fig_sim1}.a) are displayed in Figs.\ref{fig_sim2}.a-b, respectively, as a function of barrier height $U_b$ and incident angle $\theta$. Without the tilt \cite{kats06,alla11,hill17}, the transmission exhibits two well-known phenomena: $\mathcal{T}_{W,W'} = 1$ at $\theta = 0^\circ$, independent of $U_b$ (i.e., Klein tunneling), and resonant peaks in oblique directions (i.e., Fabry-P\'{e}rot interferences). Under the tilting effects, a novel feature is observed: the Klein tunneling is still preserved but is achieved in two separated oblique directions $\theta_{K}$ for the DW fermions of opposite chiralities. In particular, these Klein tunneling directions are determined by group velocities $\vec v_{K} = v_F\vec e_z + \tau_c\vec u$ with the unit vector ${\vec e}_z$ along the Oz axis, e.g., $\theta_{K} = \tau_c\cos^{-1}((v_F+\tau_c u_z)/\sqrt{(v_F+\tau_c u_z)^2 + u_\perp^2}) \simeq \pm 26.6^\circ$ for $u = 0.5 v_F$ considered in Figs.\ref{fig_sim2}.a-b. 
Note that if $\vec u \parallel$ Oz axis (i.e., $u_\perp = 0$), all transport phenomena induced by the tilting effects disappear, i.e., the similar pictures as in the non-tilted case \cite{hill17} are observed.

To clarify these results, the transmission probability $\mathcal{T}$ was computed analytically in the ideal case of abrupt barriers using matching of the wave functions at the barrier interfaces (see the details in \cite{SM2017}). In such the case, its momentum dependence is given by
\begin{eqnarray}
    \mathcal{T} = \left| f(\vec k_n,\vec k_p,0)/f(\vec k_n,\vec k_p,L_b) \right|^2
\end{eqnarray}
where $f(\vec k_n,\vec k_p,L_b) = \sin\frac{\theta^+_{\sigma n} - \theta^-_{\sigma p}}{2} \sin\frac{\theta^-_{\sigma n} - \theta^+_{\sigma p}}{2} e^{-ik^+_{pz}L_b} - \sin\frac{\theta^+_{\sigma n} - \theta^+_{\sigma p}}{2} \sin\frac{\theta^-_{\sigma n} - \theta^-_{\sigma p}}{2} e^{-ik^-_{pz}L_b}$. Here, the subscripts $n$/$p$ indicate the quantities in $n$-doped/$p$-doped zones (see Fig.\ref{fig_sim1}.a), angles $\theta^\pm_\sigma$ (vectors $\vec k^\pm$) determine the directions of pseudospin (momenta) in the transmitted and reflected states, respectively, and $L_b$ is the barrier width. %Note that $\vec k_{n,p}$ in Eq.(2) satisfy the $\vec k_\perp$-conservation \cite{SM2017}. 
It has been well known that the Klein tunneling is observed, satisfying the pseudospin conservation \cite{kats06}. With ${\vec \sigma}_{\tau_c\tau_b} = \tau_c\tau_b \vec k/k$ and the conservation of $\vec k_\perp$, the conservation $\vec \sigma^+_{n} \equiv \vec \sigma^+_{p}$ is actually obtained only for ${\vec k}_\perp = 0$ (see Fig.\ref{fig_sim1}.b) and hence the second term of $f(\vec k_n,\vec k_p,L_b)$ vanishes ($\theta^+_{\sigma n} = \theta^+_{\sigma p}$). Accordingly, the transmission $\mathcal{T} = 1$ is achieved, independent of $U_b$ and $L_b$, which demonstrates the Klein tunneling effect. The corresponding group velocity $\vec v_{K} = v_F\vec e_z + \tau_c\vec u$ (same in differently doped zones) is exactly the one reported above. Thus, the origin of the deflected Klein tunneling is simply the non-collinearity of $\vec v$ and $\vec k$ induced by the tilting effects.

In Fig.\ref{fig_sim2}.c, the dependence of the Klein tunneling direction (i.e., direction of $\vec v_{K}$) on the tilt velocity $\vec u$ is presented. While $\theta_{K}$ is a function of both $u$ and $\theta_u$, the angle $\phi_{K}$ is simply determined by $\phi_{K} = \phi_u + (1-\tau_c) 90^\circ$. Note that the direction ($\theta_{K},\phi_{K}+180^\circ$) for $\tau_c = -1$ is actually equivalent to ($-\theta_{K},\phi_{K}$) in Fig.\ref{fig_sim2}.b. As presented in Fig.\ref{fig_sim2}.c, a large deflection of the Klein tunneling is predicted when increasing the tilt velocity, i.e., $\theta_{K}$ can reach maximum values of $\sim$30$^\circ$ and $\sim$50$^\circ$ for $u = 0.5 v_F$ and $0.75v_F$, respectively. Interestingly, except for $\vec u \parallel$ Oz axis, the significant effects can be obtained in a wide range of $\theta_u$, especially when the tilt magnitude is large. 

In addition, Figs.\ref{fig_sim2}.a-b demonstrate that the Fabry-P\'{e}rot resonant spectra can be also significantly deflected when the energy dispersion is tilted. This deflection is more clearly observed in Fig.\ref{fig_sim2}.d where the transmission probability for the $W$ cone is illustrated as a function of incident angles $\theta$ and $\phi$. Indeed, instead of being isotropic with respect to the angle $\phi$ \cite{hill17} for a non-tilted dispersion, the resonant spectra become strongly anisotropic under the tilting effects, i.e., depending on both angles $\theta$ and $\phi$. Besides its dependence on the non-collinearity of $\vec v$ and $\vec k$, the deflection of Fabry-P\'{e}rot resonances can be basically explained as a consequence of the momentum shift of the Fermi surfaces as illustrated in Fig.\ref{fig_sim1}.b. Indeed, this shift, together with the conservation of $\vec k_\perp$, allows for finite transmission only in certain directions. The similar picture is achieved for the $W'$ cone, except that the resonant peaks occur in opposite directions.

Note additionally that since the non-collinearity of $\vec v$ and $\vec k$ is one of the key elements, the deflection of highly transparent directions (i.e., Klein tunneling and Fabry-P\'{e}rot resonances) can also be obtained, thank to the anisotropic character of energy dispersion (see the demonstration in \cite{SM2017}), i.e., when $v_{Fx,Fy,Fz}$ are not identical. Different from the tilting effects, this anisotropy however does not include a momentum shift of the Fermi surfaces and results in the same deflection for fermions of opposite chiralities. In general, the deflection of transparent directions reported here can be induced by the combined effects of these two characters of DW fermions. At last, it is worth noting that in addition to the pseudospin conservation, the Klein tunneling in all cases is found to exhibit another inherent property: refractionless character, i.e., the group velocity is also conserved.

\textbf{Valleytronics and electron optics in \textit{p-n} junctions} - In analogy with the valley indexes in graphene \cite{xiao07}, the two chiralities of DW fermions could be used as new degrees of freedom to store and carry information in valleytronic devices \cite{shen15,qma017}. The separated Klein tunneling directions of opposite chiralities observed in Fig.\ref{fig_sim2} suggest that valley (i.e., chirality) filtering and beam splitting effects can be obtained here. Simultaneously, the electron optics behaviors \cite{hill17} in DW \textit{p-n} junctions can be drastically modified by the tilting effects. 

\begin{figure}[!b]
	\centering
	\includegraphics[width = 0.49\textwidth]{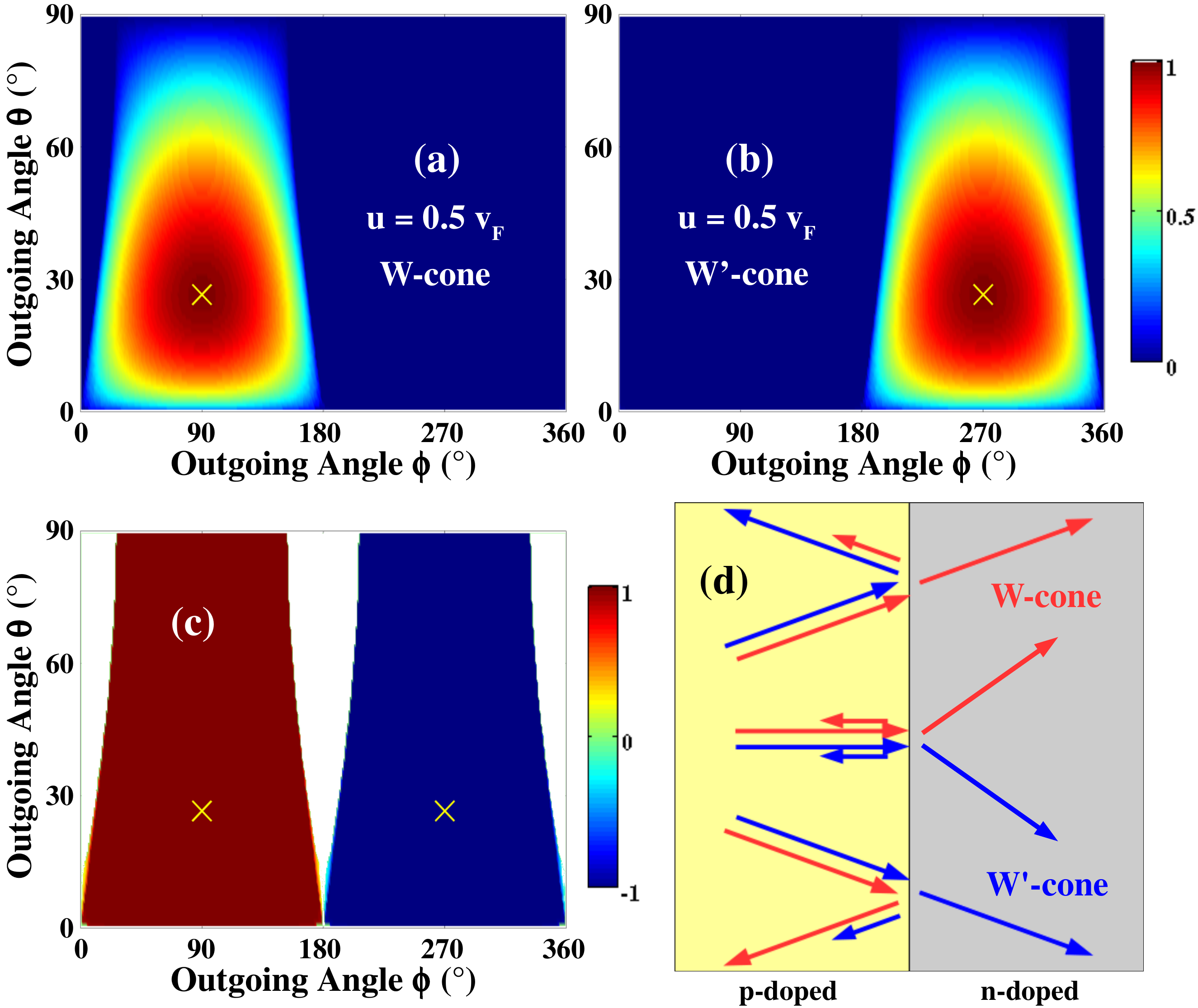}
	\caption{\textbf{Valley filtering and beam splitting.} ($\phi$-$\theta$)-maps of transmission probabilities $\mathcal{T}_{W,W'}$ (\textbf{a,b}) in a $p$-$n$ junction and the corresponding valley polarization (\textbf{c}). The the blank regions in (\textbf{c}) correspond to the fully reflective directions. The barrier height $U_b = 0.2$ while other parameters are the same as in Fig.\ref{fig_sim2}. The cross symbols mark the Klein tunneling directions. (\textbf{d}) Schematics of directional separation and beam splitting effects of valley-dependent currents.}
	\label{fig_sim4}
\end{figure}
In Fig.\ref{fig_sim4}, the transmission probabilities $\mathcal{T}_{W,W'}$ through a \textit{p-n} junction and the corresponding valley polarization $P_{val} = (\mathcal{T}_W - \mathcal{T}_{W'})/(\mathcal{T}_W + \mathcal{T}_{W'})$ are computed and presented as a function of outgoing angles $\theta$ and $\phi$. First, due to the momentum shift of the Fermi surfaces, the transmission of each DW cone is allowed only in certain directions (as already mentioned) and these transparent directions for two cones can be totally separated if the tilt is sufficiently large (see Figs.\ref{fig_sim4}.a-b). Consequently, strong valley filtering with perfect valley polarization can be obtained as depicted in Fig.\ref{fig_sim4}.c and schematically described in Fig.\ref{fig_sim4}.d. Moreover, due to the deflected Klein tunneling, the transmission probability exhibits peaks at two separated directions for the two cones (see the marked points in Figs.\ref{fig_sim4}.a-c). Hence, the high valley-dependent currents are always obtained around these two directions, independently on the barrier height and carrier energy. In addition, the high reflection directions of fermions with opposite chiralities are also separated, implying that both the transmitted and reflected beams can be highly valley polarized, as schematically described in Fig.\ref{fig_sim4}.d. 

The beam slitting effect (see also Fig.4) is most obviously demonstrated when considering the normal incident beam, i.e., $\vec v_\perp = 0$ (or $\vec k_\perp = \tau_c k_p \vec u_\perp/v_F$) in the $p$-doped zone. In such a case, $\vec v_\perp = \tau_c (1 + k_p/k_n) \vec u_\perp$ in the $n-$doped zone is obtained. Indeed, the outgoing beam is split with opposite group velocities $\vec v_\perp$ for fermions of opposite chiralities (see Fig.\ref{fig_sim4}.d). Interestingly, this effect depends not only on $\vec u_\perp$ but also on the ratio $k_p/k_n$, which is tunable by controlling the doping profile.
\begin{figure}[!b]
	\centering
	\includegraphics[width = 0.49\textwidth]{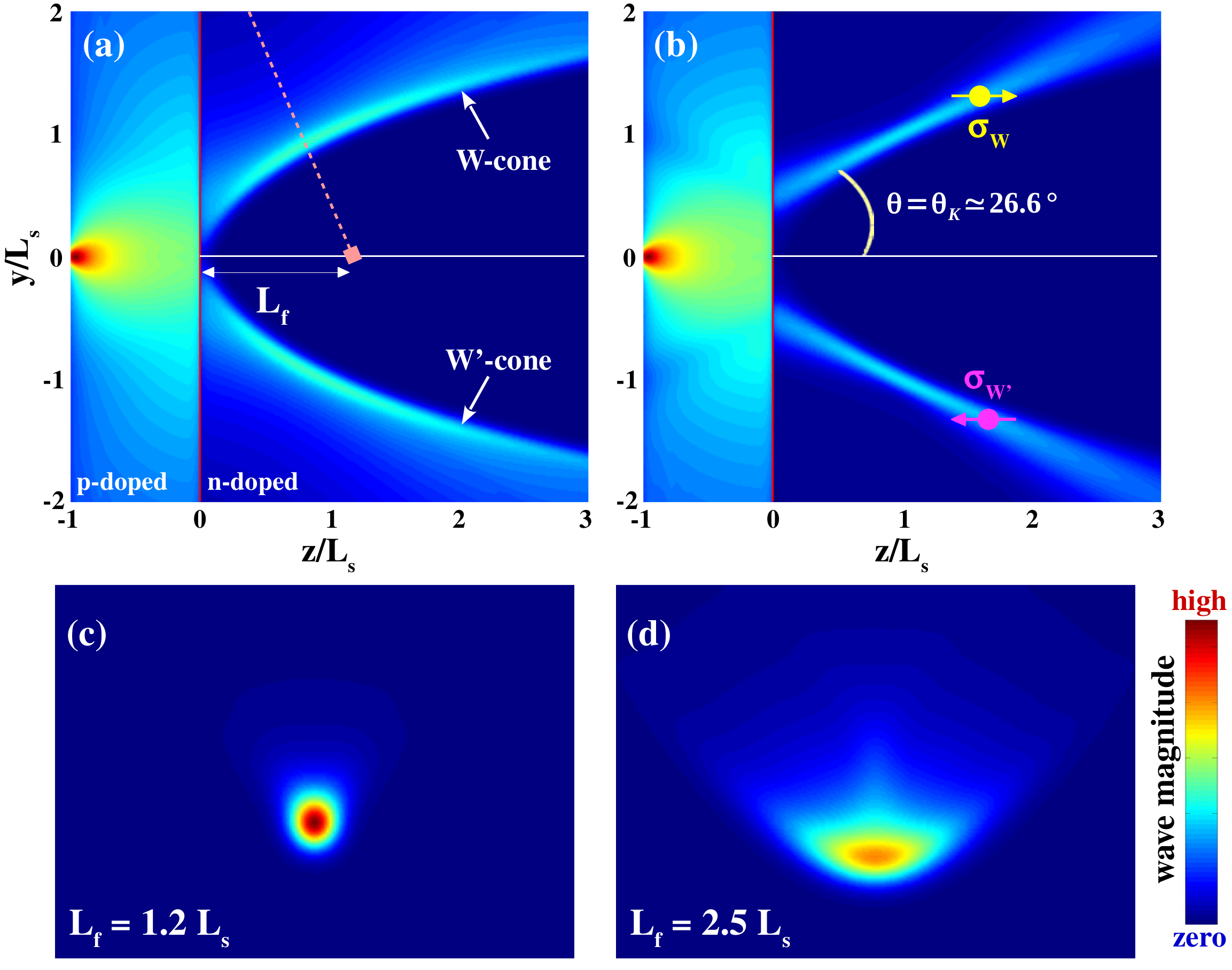}
	\caption{\textbf{Electron wave propagation in \textit{p-n} junctions} with short (\textbf{a}) and long (\textbf{b}) gradient widths of the doping profile: $L_b = 100a_0$ and 900$a_0$, respectively. Electron waves are injected from a source point at (0,0,-$L_s$) with a distance $L_s$ ($L_s \gg L_b$) from the \textit{p-n} interface. (\textbf{c,d}) represent the cut planes in (\textbf{a}), which are parallel to the Ox axis and perpendicular to the Klein tunneling direction of W-cone (see the dashed line in (\textbf{a})), with two different distances $L_f$. Here, $E = 0.1$, $U_b = 0.2$, and $\vec u$ = $u(0,1,0)$ with $u = 0.5 v_F$.}
	\label{fig_sim5}
\end{figure}

At last, the tilting effects on the optical-like behaviors of DW fermions in \textit{p-n} junctions should be clarified. The refraction of electron waves at the \textit{p-n} interface in the isotropic dispersion case is known to satisfy the Snell's law with negative refractive indexes, inducing the Veselago focusing effect \cite{alla11,chei07,wilm14,hlee15,hill17}. This picture can be dramatically disturbed by the tilting effects (see Fig.\ref{fig_sim5}). To mimic the electron optics behaviors, the propagation of electron waves presented in Fig.\ref{fig_sim5} was computed using a semi-classical billiard model (see Ref. \cite{SM2017} for more details). Actually, starting from the Snell's law, the tilting effects are found to split the outgoing waves of opposite chiralities into opposite directions (see Figs.\ref{fig_sim5}.a-b), thus inducing both the positive (around the Klein tunneling directions) and negative refraction indexes \cite{SM2017}. As an amazing result, the tilted DW system exhibits two focusing points for opposite chiralities as displayed in Figs.\ref{fig_sim5}.a and \ref{fig_sim5}.c-d whereas a unique point is observed in the normal direction of the isotropic fermion systems \cite{chei07,hill17,SM2017}. 

In addition, while all the other transmissions can be strongly affected, the Klein tunneling is a unique one that is insensitive to an increase of gradient width between highly doped zones \cite{alla11}. This property has been demonstrated to be potential for designing Klein collimators \cite{hliu17} when the gradient width is long enough. In such case, the tilted DW systems represent two separated oblique collimation directions for opposite chiralities as clearly visible in Fig.\ref{fig_sim5}.b. Moreover, besides owning to two opposite chiralities, these two Klein collimated beams also possess well-defined and anti-parallel pseudospins, i.e., ${\vec \sigma}_{\tau_c} = \tau_c \vec \xi_{K}$, thus offering promising possibilities in both valleytronics and pseudospintronics.

\textbf{Discussion and conclusion} - Experimental confirmation of phenomena explored in this work requires the realization and controllability of the system doping profile. In this regard, the electrostatic doping could be considered as the most appropriate technique. Note that our findings are valid in both 2D and 3D DW systems. While it has been demonstrated in 2D cases, the electrostatic doping in 3D systems may require some further developments, in particular, to solve the issue of gate efficiency. In the latter case, some directions have already been demonstrated, i.e., the gate doping can be efficiently applied in DW thin films \cite{yliu15} and nanowires \cite{czli15}. However, another issue, i.e., finite size-induced confinements, could arise and should be taken into account. In particular, a crossover from 3D to 2D electronic behaviors was observed in \cite{xxia17} when the thickness is reduced below 26nm while the characteristic of Dirac particles was still observed in \cite{hell17} with a film thickness of 18nm. In addition, the directionally separated currents predicted can be measured using multiple directional leads as in \cite{hlee15,chen16}.

To conclude, with tilted energy dispersions, DW fermions possess some exceptional properties: (i) non-collinearity of $\vec v$ and $\vec k$ (similarly, $\vec v$ and pseudospin $\vec \sigma$); and (ii) momentum shift of the Fermi surface when varying the carrier energy. These properties dramatically modify their transport properties when compare to the isotropic dispersion case. In particular, the Klein tunneling of tilted fermions of opposite chiralities is observed in two separated oblique directions, instead of occurring in a unique  one (i.e., normal incidence) as in the non-tilted case. In addition, all electron optics behaviors in the hetero-doped junctions can be strongly modified by the tilt, thus inducing possible valley filtering and beam splitting effects. In summary, the present study highlights the outstanding transport properties of tilted DW fermions, that could pave the way for novel applications of the host materials in electron optics and valleytronics.

V.H.N. and J.-C.C. acknowledge financial support from the F.R.S.-FNRS of Belgium through the research project (N$^\circ$ T.1077.15), from the F\'{e}d\'{e}ration Wallonie-Bruxelles through the ARC on 3D architecturing of 2D crystals (N$^\circ$ 16/21-077) and from the European Union's Horizon 2020 research and innovation program (N$^\circ$ 696656).

\clearpage

\newpage

\setcounter{equation}{0}
\renewcommand\theequation{S\arabic{equation}}

\setcounter{figure}{0}
\renewcommand\thefigure{S\arabic{figure}}

\setcounter{table}{0}
\setcounter{page}{1}

\onecolumngrid

\makeatletter

	\begin{center}
		{\Large \textbf{Supplemental Material for}} \\ {\large "Klein tunneling and electron optics in Dirac-Weyl fermion systems \\ with tilted energy dispersion"}
	\end{center}

\subsection{A. Green's function calculations}

The Dirac-Weyl Hamiltonian is written in the Letter in the following form:
\begin{equation}
H_0 = \sum\limits_{\iota = x,y,z} \tau_c  (u_{\iota} p_{\iota} + \sigma_{\iota} v_{F\iota} p_{\iota})
\end{equation}
where $\tau_c = \pm 1$ represent the chirality, $\sigma_{\iota}$ are the Pauli's matrices and $\vec u$ describes the tilt of energy dispersion. Considering nanostructures formed by applying a potential profile $U(z)$ along the Oz axis (see Fig.1.a in the Letter), the system Hamiltonian reads $H = H_0 + U(z)$. 

In order to compute the transport properties, the calculation method developed in Refs. \cite{hung10,habi16} is employed. First, the Hamiltonian $H$ is separated into two independent parts, i.e., $H = H_z +H_\perp$ where $H_z = \tau_c (u_z p_z + v_{Fz} \sigma_z p_z) + U(z)$ and $H_\perp = \tau_c (u_x p_x + v_{Fx} \sigma_x p_x) + \tau_c (u_y p_y + v_{Fy} \sigma_y p_y)$. To solve it numerically, $H$ is then represented in the basis $\left\lbrace \left| z_m \right\rangle \otimes \left| \vec k_\perp \right\rangle \right\rbrace$ with the mesh-spacing $a_0 = z_{m+1} - z_m$ along the transport direction (i.e., Oz axis / perpendicular to the junction interfaces) and the plane wave $\left| \vec k_\perp \right\rangle = e^{i\vec k_\perp \vec r_\perp}/\sqrt{S}$, and $S$ is the system area in the Oxy plane. Finally, the Hamiltonian can be rewritten in the following form:
\begin{eqnarray} 
H &=& \sum\limits_{\vec k_\perp} H(\vec k_\perp), \\
H(\vec k_\perp) &=& \sum\limits_{m} \left[ \left\lbrace  U(z_m) + \tau_c \epsilon (\vec k_\perp) \right\rbrace \left| z_m \right\rangle \left\langle z_m \right| - \tau_c \beta \left\lbrace i \left| z_m \right\rangle \left\langle z_{m-1} \right| - i \left| z_m \right\rangle \left\langle z_{m+1} \right| \right\rbrace \right], \nonumber
\end{eqnarray}
with $\epsilon (\vec k_\perp) = \hbar (u_x + v_{Fx}\sigma_x) k_x + \hbar (u_y + v_{Fy}\sigma_y) k_y$ and $\beta = \hbar \left(u_z + v_{Fz} \sigma_z \right)/2a_0$. Using this form of the Hamiltonian, the retarded device Green's function is determined by \cite{hung10}
\begin{eqnarray} 
G_D(E,\vec k_\perp) &=& \left[ E + i0^+ - H_D(\vec k_\perp) - \Sigma_{L} (E,\vec k_\perp) - \Sigma_{R} (E,\vec k_\perp) \right]^{-1},
\end{eqnarray}
where the Hamitonian $H_D$ models the active region and $\Sigma_{L,R}$ are the left and right contact self-energies, respectively. The transport properties through the system are then extracted. In particular, the transmission probability is computed as $\mathcal{T} =$ Tr$[\Gamma_L G_D \Gamma_R G_D^\dagger]$ with $\Gamma_{L,R} = i(\Sigma_{L,R} - \Sigma_{L,R}^\dagger)$.

\subsection{B. Analytical solution for abrupt potential barriers}

In this section, the Hamiltonian (S1) is solved analytically to compute the transmission through the structures of single abrupt potential barrier. The energy eigenvalues and corresponding wavefunctions are given by
\begin{eqnarray}
E_{\tau_c\tau_b} (\vec k) &=& \hbar \left( \tau_c \vec u \vec k + \tau_b |\vec \xi_{\vec k}| \right) \\
\psi_{\tau_c\tau_b} (\vec k_\perp, k_z, \vec r) &=& \frac{1}{\sqrt{2 |\vec \xi_{\vec k}| (|\vec \xi_{\vec k}| + \tau_c\tau_b \xi_{\vec k, z})}} 
\left[ 
\begin{array}{cc}
|\vec \xi_{\vec k}| + \tau_c\tau_b \xi_{\vec k, z} \\
\tau_c\tau_b (\xi_{\vec k, x} + i \xi_{\vec k, y})
\end{array} 
\right] e^{ik_z z} e^{i\vec k_\perp \vec r_\perp}
\end{eqnarray}
where $\vec \xi_{\vec k} = (v_{Fx} k_x, v_{Fy} k_y, v_{Fz} k_z)$ and $\tau_b = \pm 1$ corresponding to the conduction/valence bands, respectively. Accordingly, the pseudo-spin $\vec \sigma$ and group velocity $\vec v$ of this fermion system are determined as
\begin{eqnarray}
{\vec \sigma}_{\tau_c\tau_b} (\vec k) = \tau_c \tau_b \vec \xi_{\vec k} / |\vec \xi_{\vec k}| \,\,\,\, \mathrm{and} \,\,\,\, {\vec v}_{\tau_c\tau_b} (\vec k) = \tau_c \vec u + \tau_b \vec \eta_{\vec k} / |\vec \xi_{\vec k}|,
\end{eqnarray}
with $\vec \eta_{\vec k} = (v_{Fx}^2 k_x, v_{Fy}^2 k_y, v_{Fz}^2 k_z)$. The wavefunctions in the equation (S5) can be also rewritten in another simple form:
\begin{equation}
\psi_{\tau_c\tau_b} (\vec k_\perp, k_z, \vec r) = 	\left[ 
\begin{array}{cc}
\cos\frac{\theta_\sigma}{2} \\
\sin\frac{\theta_\sigma}{2} e^{i\phi_\sigma}
\end{array} 
\right] e^{ik_z z} e^{i\vec k_\perp \vec r_\perp}
\end{equation}
Here, the angles $\theta_\sigma$ and $\phi_\sigma$ determine the direction of $\vec \sigma$ in the equation (S6), i.e., $\vec \sigma = (\sin\theta_\sigma\cos\phi_\sigma,\sin\theta_\sigma\sin\phi_\sigma,\cos\theta_\sigma)$.

The velocities $v_{F\iota}$ in the equation (S1) are now assumed to be isotropic, i.e., $v_{Fx} \equiv v_{Fy} \equiv v_{Fz} = v_F$, and hence ${\vec \sigma}_{\tau_c\tau_b} = \tau_c\tau_b \vec k/k$, i.e., $\theta_\sigma = \theta_k - \theta_c$ and $\phi_\sigma \equiv \phi_k$ with $\vec k = k(\sin\theta_k\cos\phi_k,\sin\theta_k\sin\phi_k,\cos\theta_k)$ and $\theta_c = (1 - \tau_c \tau_b) \pi/2$. The wavefunctions in three (left-outside, inside, right-outside) zones of a single potential barrier system (see Fig.1.a in the Letter) can be respectively written in the following forms:
\begin{eqnarray}
\psi_1 &=& \left[ \begin{array}{cc} \cos\frac{\theta^+_{\sigma n}}{2} \\ \sin\frac{\theta^+_{\sigma n}}{2} e^{i\phi_\sigma} \end{array} \right] e^{izk^+_{nz}} e^{i\vec k_\perp \vec r_\perp} + r\left[ \begin{array}{cc} \cos\frac{\theta^-_{\sigma n}}{2} \\ \sin\frac{\theta^-_{\sigma n}}{2} e^{i\phi_\sigma} \end{array} \right] e^{izk^-_{nz}} e^{i\vec k_\perp \vec r_\perp}, \\ 
\psi_2 &=& a\left[ \begin{array}{cc} \cos\frac{\theta^+_{\sigma p}}{2} \\ \sin\frac{\theta^+_{\sigma p}}{2} e^{i\phi_\sigma} \end{array} \right] e^{izk^+_{pz}} e^{i\vec k_\perp \vec r_\perp} + b \left[ \begin{array}{cc} \cos\frac{\theta^-_{\sigma p}}{2} \\ \sin\frac{\theta^-_{\sigma p}}{2} e^{i\phi_\sigma} \end{array} \right] e^{izk^-_{pz}} e^{i\vec k_\perp \vec r_\perp}, \\
\psi_3 &=&  t\left[ \begin{array}{cc} \cos\frac{\theta^+_{\sigma n}}{2} \\ \sin\frac{\theta^+_{\sigma n}}{2} e^{i\phi_\sigma} \end{array} \right] e^{izk^+_{nz}} e^{i\vec k_\perp \vec r_\perp},
\end{eqnarray}
where the superscripts +/- indicate the transmitted/reflected states while the momenta $k^+_{pz}$ (and $k^-_{pz}$) and $k^+_{nz}$ (and $k^-_{nz}$) satisfy the equations:
\begin{eqnarray}
&&\hbar \tau_c (\vec u_\perp \vec k_\perp + u_z k_{pz}) - \hbar v_F \sqrt{k_\perp^2 + k_{pz}^2} = E - U_b \\
&&\hbar \tau_c (\vec u_\perp \vec k_\perp + u_z k_{nz}) + \hbar v_F \sqrt{k_\perp^2 + k_{nz}^2} = E
\end{eqnarray}
respectively. Using the continuity of the wave function at $z = 0$ and $z = L_b$, the coefficient $t$ in the equation (S10) is given by
\begin{equation}
t = \frac{\sin\frac{\theta^+_{\sigma n} - \theta^-_{\sigma p}}{2} \sin\frac{\theta^-_{\sigma n} - \theta^+_{\sigma p}}{2} - \sin\frac{\theta^+_{\sigma n} - \theta^+_{\sigma p}}{2} \sin\frac{\theta^-_{\sigma n} - \theta^-_{\sigma p}}{2}} {\sin\frac{\theta^+_{\sigma n} - \theta^-_{\sigma p}}{2} \sin\frac{\theta^-_{\sigma n} - \theta^+_{\sigma p}}{2} e^{-ik^+_{pz}L_b} - \sin\frac{\theta^+_{\sigma n} - \theta^+_{\sigma p}}{2} \sin\frac{\theta^-_{\sigma n} - \theta^-_{\sigma p}}{2} e^{-ik^-_{pz}L_b}} e^{-ik^+_{nz}L_b}
\end{equation}
The transmission probability is finally determined as
\begin{eqnarray}
\mathcal{T} = \left| t \right|^2 = \frac{\left( \sin\frac{\theta^+_{\sigma n} - \theta^-_{\sigma p}}{2} \sin\frac{\theta^-_{\sigma n} - \theta^+_{\sigma p}}{2} - \sin\frac{\theta^+_{\sigma n} - \theta^+_{\sigma p}}{2} \sin\frac{\theta^-_{\sigma n} - \theta^-_{\sigma p}}{2} \right)^2 } {\left| \sin\frac{\theta^+_{\sigma n} - \theta^-_{\sigma p}}{2} \sin\frac{\theta^-_{\sigma n} - \theta^+_{\sigma p}}{2} e^{-ik^+_{pz}L_b} - \sin\frac{\theta^+_{\sigma n} - \theta^+_{\sigma p}}{2} \sin\frac{\theta^-_{\sigma n} - \theta^-_{\sigma p}}{2} e^{-ik^-_{pz}L_b} \right|^2 },
\end{eqnarray}
which is exactly the equation (2) for $\mathcal{T}$ presented in the Letter. 

\subsection{C. Refraction and reflection of electron waves at \textit{p-n} interface}

The tilting effects on the refraction and reflection of electron waves at a \textit{p-n} interface are discussed in this section. As mentioned, the momentum $k_z$ in the \textit{p}- and \textit{n}-doped zones satisfies the equations (S11) and (S12), respectively. %Here, the ${\vec k}_\perp$ - conservation is already taken into account. 
For simplicity, energies $E_n \equiv E$ and $E_p \equiv U_b - E$ will be used below as two energy parameters. Note that both equations (S11) and (S12) have two possible solutions $k^+_z$ (transmitted) and $k^-_z$ (reflected waves). The group velocities in these two doped zones are respectively given by
\begin{eqnarray}
{\vec v}_p = \tau_c \vec u - v_F {\vec k}_p /k_p  \,\,\,\, \mathrm{and} \,\,\,\, {\vec v}_n = \tau_c \vec u + v_F {\vec k}_n /k_n
\end{eqnarray} 
The momentum $k^+_z$ (resp. $k^-_z$) is determined by the condition $v^+_z > 0$ (resp. $v^-_z < 0$). The vectors ${\vec v}_p$ and ${\vec v}_n$ are then rewritten in the form $\vec v = v (\sin\theta\cos\phi,\sin\theta\sin\phi,\cos\theta)$. Solving the equations (S11,S12,S15), the propagation/reflection directions of an electron wave and its refraction at the $p-n$ interface are determined.

Let us consider a simple case when $\vec u$ is parallel to the Oy axis, i.e., $\vec u = u (0,1,0)$. First, the incoming electron waves are assumed to propagate only in the Oyz plane in the \textit{p-}doped zone, i.e., $v_{px} =0$. Accordingly, $k_x = 0$ and hence $v_{nx} = 0$ is similarly obtained for the outgoing waves in the \textit{n-}doped zone. The group velocities in these two doped zones can hence be written as $\vec v_p = v_p(0,\sin\theta_p,\cos\theta_p)$ and $\vec v_n = v_n(0,\sin\theta_n,\cos\theta_n)$.
\begin{figure}[!t]
	%	\centering
	\includegraphics[width = 0.99\textwidth]{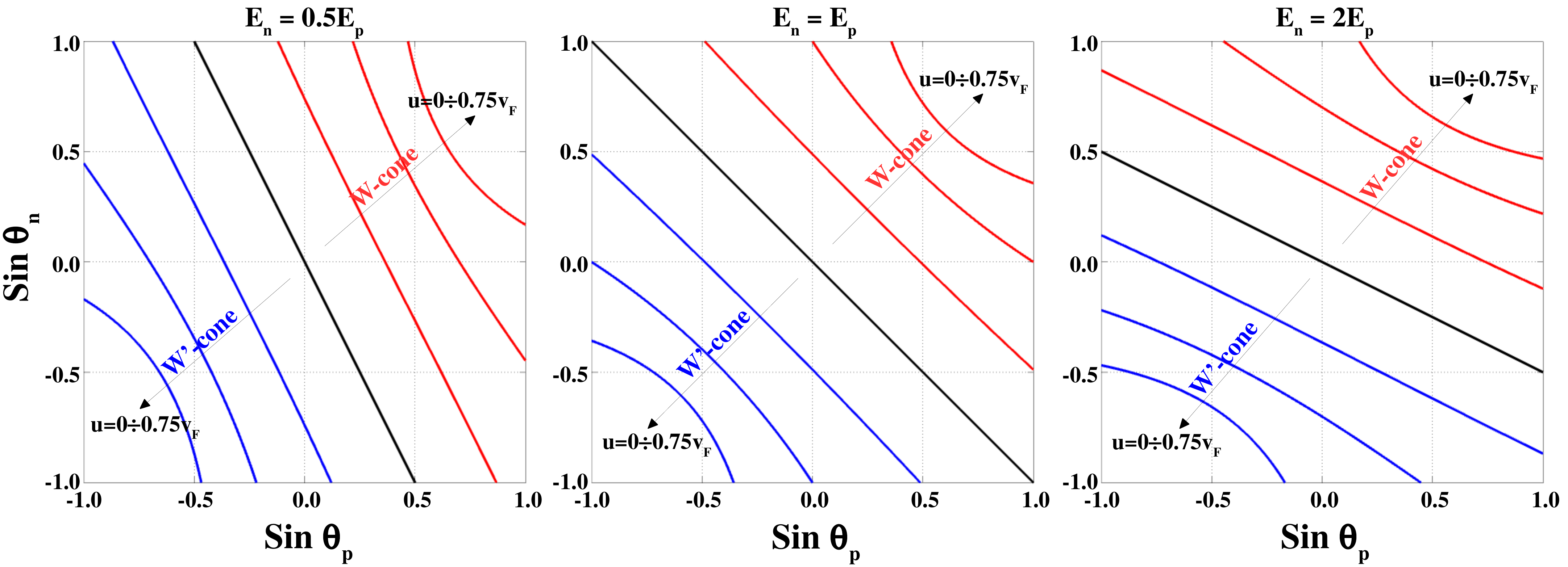}
	\caption{\textbf{Refraction of electron waves at a $p-n$ interface under the tilting effects.} The incoming waves are assumed to propagate in the Oyz plane in the $p-$doped zone. The relationship between incoming $\theta_p$ and outgoing $\theta_n$ angles with different tilt velocities and potential barriers $U_b$ (i.e., $U_b = E_n + E_p$) is presented. All calculations are performed with $E_p = 0.1$ and $(\theta_u,\phi_u) = (90^\circ,90^\circ)$.}
	\label{fig_sim3}
\end{figure}
Solving the equations (S11,S12,S15), the refraction rule describing the relationship between propagation angles $\theta_p$ and $\theta_n$ is obtained and displayed in Fig.\ref{fig_sim3}. Actually, the Snell's law obtained in the non-tilted case satisfies the simple equation $E_n\sin\theta_n = -E_p\sin\theta_p$ (in another equivalent form, $k_n\sin\theta_n = -k_p\sin\theta_p$) with negative refraction indexes. Under the tilting effects as presented in Fig.\ref{fig_sim3}, the refraction rule can be strongly modified, in particular, both negative and positive refraction indexes, together with the directional separation of electron waves of opposite chiralities, can be observed. For a simple description, the new refraction rule obtained for small tilts (i.e., $u < 0.5 v_F$) can be approximately expressed as
\begin{equation}
E_n\sin\theta_n + E_p\sin\theta_p = \tau_c\frac{4u}{v_F} \frac{E_n E_p}{E_n+E_p}
\end{equation}
%This is obviously a different refraction rule, compared to the the Snell's law of isotropic fermions.

We now investigate the incoming electron waves propagating in the Oxz plane in the $p$-doped zone, i.e., $v_{py} = 0$. Accordingly, $k_y = \tau_c \frac{E_p}{\hbar v_F} \frac{u v_F}{v_F^2 - u^2}$, $k_p = \frac{E_p}{\hbar v_F} \frac{v_F^2}{v_F^2 - u^2}$ and $k_n = \frac{v_F^2 E_n- u^2 (E_n + E_p)}{\hbar v_F (v_F^2 - u^2)}$ are obtained.
\begin{figure}[!b]
	%	\centering
	\includegraphics[width = 0.5\textwidth]{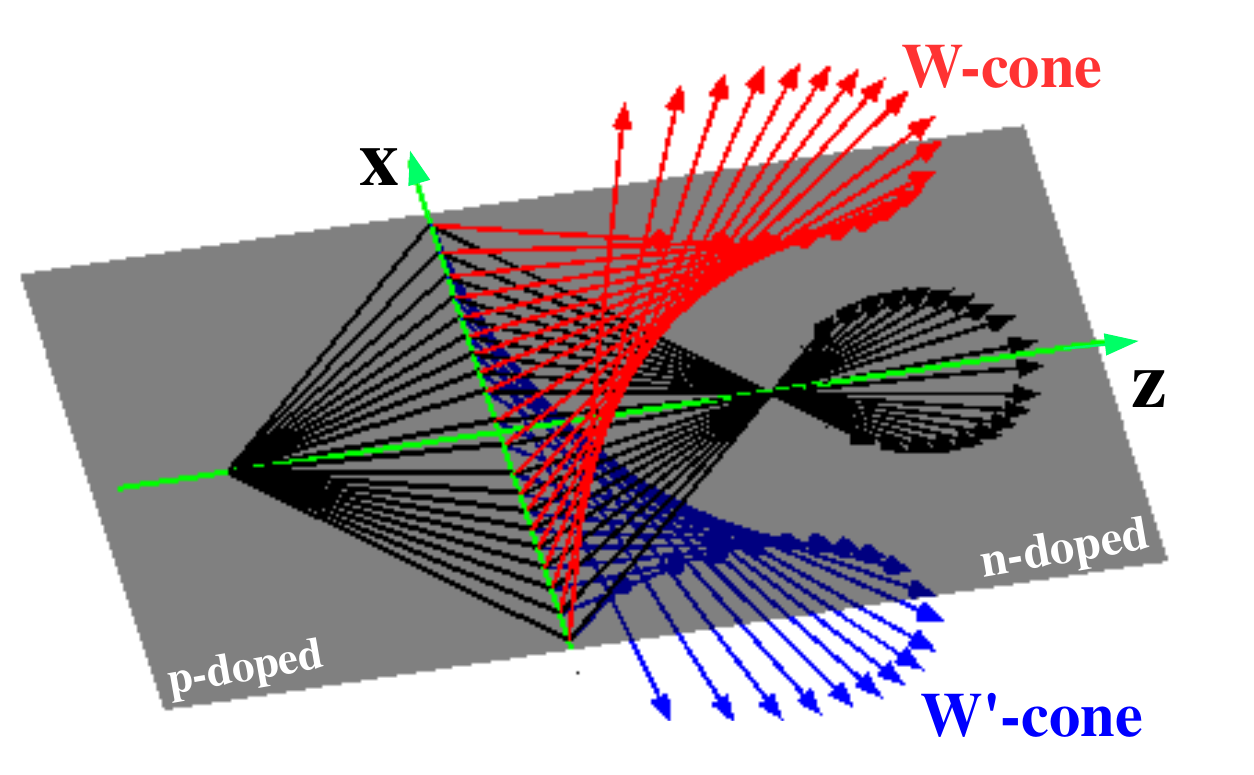}
	\caption{\textbf{Refraction of electron waves at a $p-n$ interface under the tilting effects.} The incoming waves are assumed to propagate in the Oxz plane in the $p-$doped zone and their ballistic trajectories are presented. In the $n-$doped zone, the black arrows describe the trajectory of electrons with an isotropic energy dispersion and the rest is obtained for $u = 0.3 v_F$ and $(\theta_u,\phi_u) = (90^\circ,90^\circ)$ while $E_n = E_p = 0.1$ in both cases.}
	\label{fig_sim6}
\end{figure}
The group velocities in two doped zones are then determined as follows:
\begin{eqnarray}
&&v_{px} = - v_F \frac{k_x}{k_p}, \,\,\,\,\,\,\,\,\,\,\,\,\,\,\,\,\,\,\,\,\,\,\,\,\, v_{nx} = v_F \frac{k_x}{k_n} \nonumber \\  
&&v_{py} = 0, \,\,\,\,\,\,\,\,\,\,\,\,\,\,\,\,\,\,\,\,\,\,\,\,\,\,\,\,\,\,\,\,\,\,\,\,\,\,\,\, v_{ny} = \tau_c u \frac{(v_F^2-u^2)(E_n + E_p)}{v_F^2 E_n- u^2 (E_n + E_p)} \\  
&&v_{pz} = - v_F \frac{k_{pz}}{k_p}, \,\,\,\,\,\,\,\,\,\,\,\,\,\,\,\,\,\,\,\,\,\,\, v_{nz} = v_F \frac{k_{nz}}{k_n} \nonumber
\end{eqnarray}
Here, two additional peculiar features, compared to the non-tilted case, are found. Indeed, while the momentum $k_p$ is a function of only $E_p$, the corresponding values of $k_n$ allowing for finite transmission depend on both energies $E_p$ and $E_n$. More astonishingly, the outgoing waves of opposite chiralities in the $n$-doped zone escape in opposite directions (i.e., $v_{ny}(\tau_c=1) = - v_{ny}(\tau_c=-1) \neq 0$) and hence no longer travel in the same plane with the incoming ones, as demonstrated by the equation (S17) and in Fig.\ref{fig_sim6}. However, considering the projection of electron waves in the Oxz plane, the fully negative refraction is likely obtained, i.e., $v_{px}v_{nx} = -\frac{k_x^2}{k_nk_p}v_F^2 < 0$ while both $v_{pz}$ and $v_{nz} > 0$. 

Considering all the propagation directions of incoming waves, we find an overall conclusion that starting from the Snell's law, the tilting effects tend to split the outgoing waves of opposite chiralities into opposite directions and thus induce both the positive and negative refraction indexes. Note that since, as discussed in the Letter, the Klein tunneling is a refractionless process, the wave propagation in the directions around the Klein tunneling one always undergoes a positive refraction. As an important consequence (see the discussions for Fig.4 in the Letter), two Veselago focusing points of fermions with opposite chiralities can be obtained in two different oblique axes. Moreover, in the case if the width of gradient zone between highly doped ones is long enough, two collimated beams are correspondingly observed in two deflected Klein tunneling directions explored in the Letter.

\begin{figure}[!b]
	%	\centering
	\includegraphics[width = 0.8\textwidth]{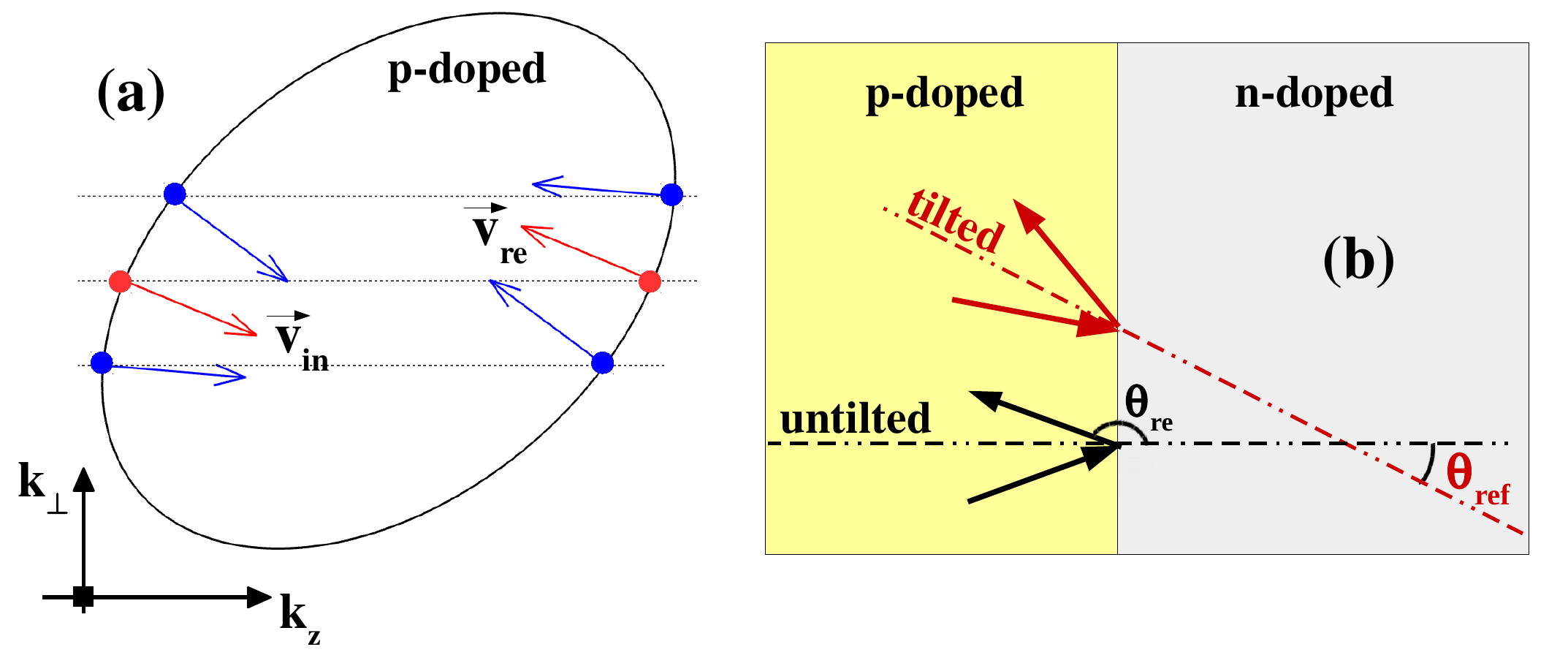}
	\caption{\textbf{Reflection of electron waves at a $p-n$ interface under the tilting effects.} Calculations in (\textbf{a}) are performed with $u = 0.75 v_F$ and $(\theta_u,\phi_u) = (40^\circ,90^\circ)$. The horizontal/black lines in (\textbf{a}) imply the ${\vec k}_\perp$ -conservation while ${\vec v}_{in}$ (${\vec v}_{re}$) determines the direction of incoming (corresponding reflected) waves.}
	\label{fig_sim7}
\end{figure}
Now, we would like to discuss some effects of tilted dispersion on the reflection of electron waves at a $p-n$ interface. In the cases of isotropic energy dispersions, it has been well known that the conventional mirror reflection with respect to the normal direction takes place and simultaneously the normal incident beam is the unique one undergoing a retro-reflection. This reflection rule however can be modified by the tilting effects. To clarify this point, we now investigate the simple case of $\vec u = (0,u_y,u_z)$ and first focus on the normal incident transmission, i.e., $v_{px}^+ = v_{py}^+ = 0$ in the $p-$doped zone. Accordingly, $k_x = 0$ and $k_y = \tau_c k_p^+ u_y/v_F$ are obtained. The solution of the equation (S11) is hence given by
\begin{eqnarray}
k_{pz}^{\pm} &=& \frac{\tau_c \hbar u_z (E_p + \tau_c \hbar u_y k_y) \mp \sqrt{E_p^2 + 2\tau_c \hbar u_y k_y E_p + \hbar^2 (u_y^2 + u_z^2 - v_F^2)k_y^2}}{\hbar^2 (v_F^2 - u_z^2)} \nonumber \\
k_{p}^{\pm} &=& \frac{\hbar^2 v_F^2 (E_p + \tau_c \hbar u_y k_y) \mp \tau_c \hbar u_z \sqrt{E_p^2 + 2\tau_c \hbar u_y k_y E_p + \hbar^2 (u_y^2 + u_z^2 - v_F^2)k_y^2}}{\hbar^3 v_F (v_F^2 - u_z^2)} \nonumber
\end{eqnarray}
Here, the superscripts +/- indicate the transmitted and reflected states, respectively. For the reflected state, $v_{px}^- = 0$ and $v_{py}^- = \tau_c u_y (1 - \frac{k_p^+}{k_p^-})$ are finally obtained. If $u_z = 0$ (consequently, $k_p^+ \equiv k_p^-$) or $u_y = 0$, $v_{py}^- = 0$ is also obtained. Thus, in such these cases, the normal incident transmission is exactly the process, in which a retro-reflection takes place, and accordingly the reflection of electron waves undergoes the conventional mirror rule. However, if $u_y u_z \neq 0$, $v_{py}^- \neq 0$ is achieved and such the conventional picture is therefore no longer valid. Indeed, as demonstrated in Fig.\ref{fig_sim7}, the retro-reflection is now observed in an oblique one $\theta_{ref}$ (see the red arrows in Fig.\ref{fig_sim7}.a). This feature is similar to that observed in a normal-metal-superconductor interface in graphene  \cite{been06}. Consequently, while the conventional one is invalid, the mirror reflection likely takes place with respect to the mentioned oblique direction $\theta_{ref}$ as illustrated in Fig.\ref{fig_sim7}.b. Moreover, this effect leads to an asymmetry between the reflected beams with angles $\theta_{re} > 180^\circ$ and $< 180^\circ$.

\subsection{D. Electron wave propagation in p-n junctions: semi-classical billiard model}
\begin{figure}[!b]
	%	\centering
	\includegraphics[width = 0.9\textwidth]{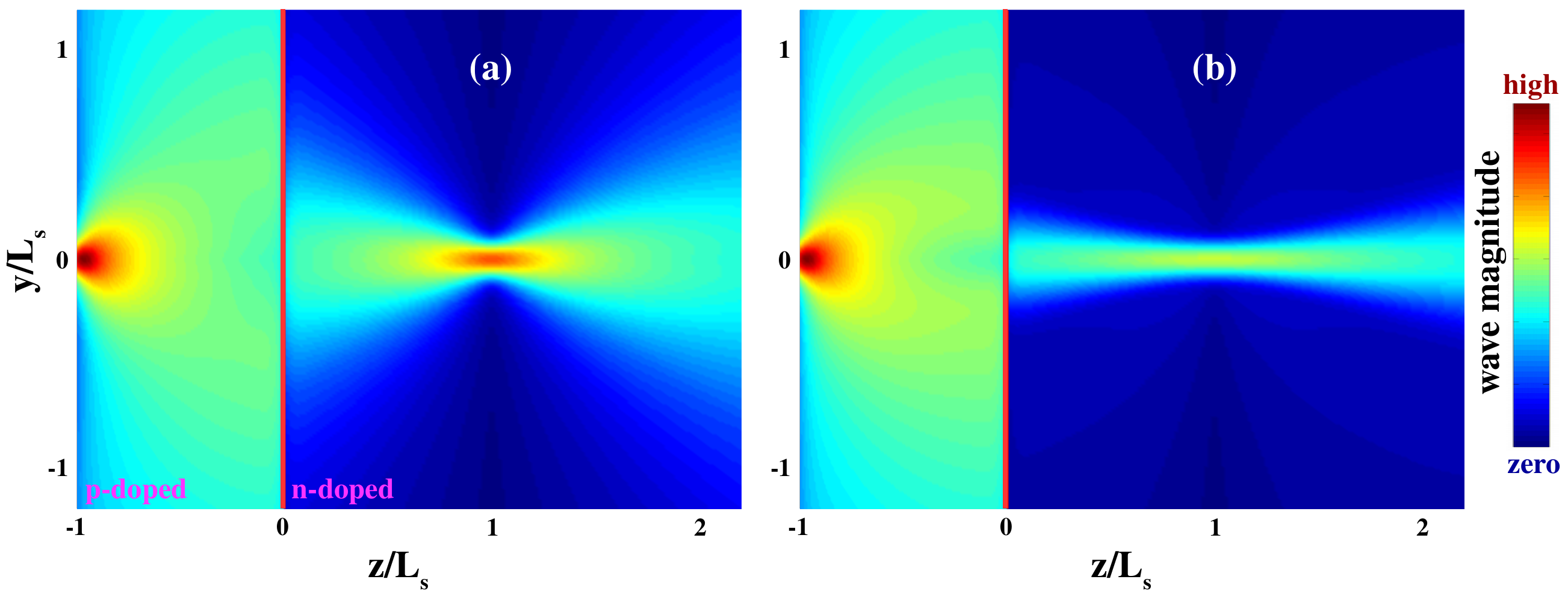}
	\caption{\textbf{Electron wave propagation in DW \textit{p-n} junctions} with short (\textbf{a}) and long (\textbf{b}) gradient widths: $L_b = 100a_0$ and 900$a_0$, respectively, in the non-tilted cases. Electron waves are injected from a source point at (0,0,-$L_s$) with a distance $L_s$ ($L_s \gg L_b$) from the \textit{p-n} interface. Calculations are performed with $E = 0.1$ and $U_b = 0.2$.}
	\label{fig_sim8}
\end{figure}

In order to clarify the effects of tilted energy dispersion on the electron optics behaviors in DW $p-n$ junctions, we compute the propagation of electron waves using a simple numerical scheme based on the semi-classical billiard model as in Refs.\cite{milo14,milo15,chen16} without a magnetic field. In this model, electrons are considered as point particles following classical/ballistic trajectories and are injected from a source point (similarly as in Refs.\cite{hliu17,libi17}) in the left doped (i.e. \textit{p-}doped) zone while the injection angle is assumed to follow a normal distribution. We investigate here the cases when the system size is much larger than the width of gradient zone between highly doped ones and hence the electron trajectories at the junction interface simply undergo the refraction/reflection and the transmission probability obtained by the calculations in the previous sections. As demonstrated in Fig.\ref{fig_sim8}, this model works very well in description of two typical pictures \cite{milo15,libi17} of isotropic DW fermions: Veselago focusing (a) and Klein collimation effects (b) for short and long gradient widths, respectively. The model was employed to compute the electron wave propagation in tilted DW $p-n$ junctions presented in Fig.4 of the Letter.

\subsection{E. The effects of anisotropic energy dispersion}

Actually, the non-collinearity of group velocity $\vec v$ and momentum $\vec k$ can be obtained not only by the tilt but also by the anisotropy of energy dispersion (see in the section B). Both these tilting and anisotropic characters have been observed in several materials as mentioned in the Letter. Due to the induced non-collinearity of  $\vec v$ and $\vec k$, it is anticipated that these two characters could have some similar effects on the transport properties of the system. To clarify this point, the tilting effects are now neglected and the effects of anisotropic energy dispersion are considered separately. Accordingly, the group velocity and pseudospin are determined by the equation (S6) with $\vec u = 0$. Here, the anisotropic dispersion is simply characterized by unidentical velocities $v_{Fx}$ $v_{Fy}$ and $v_{Fz}$ in three symmetry directions. Except if $\vec k$ is parallel to one of these symmetry directions,  $\vec v$ and $\vec k$ are non-collinear under the effects of this anisotropic character.
\begin{figure}[!b]
	\centering
	\includegraphics[width = 0.88\textwidth]{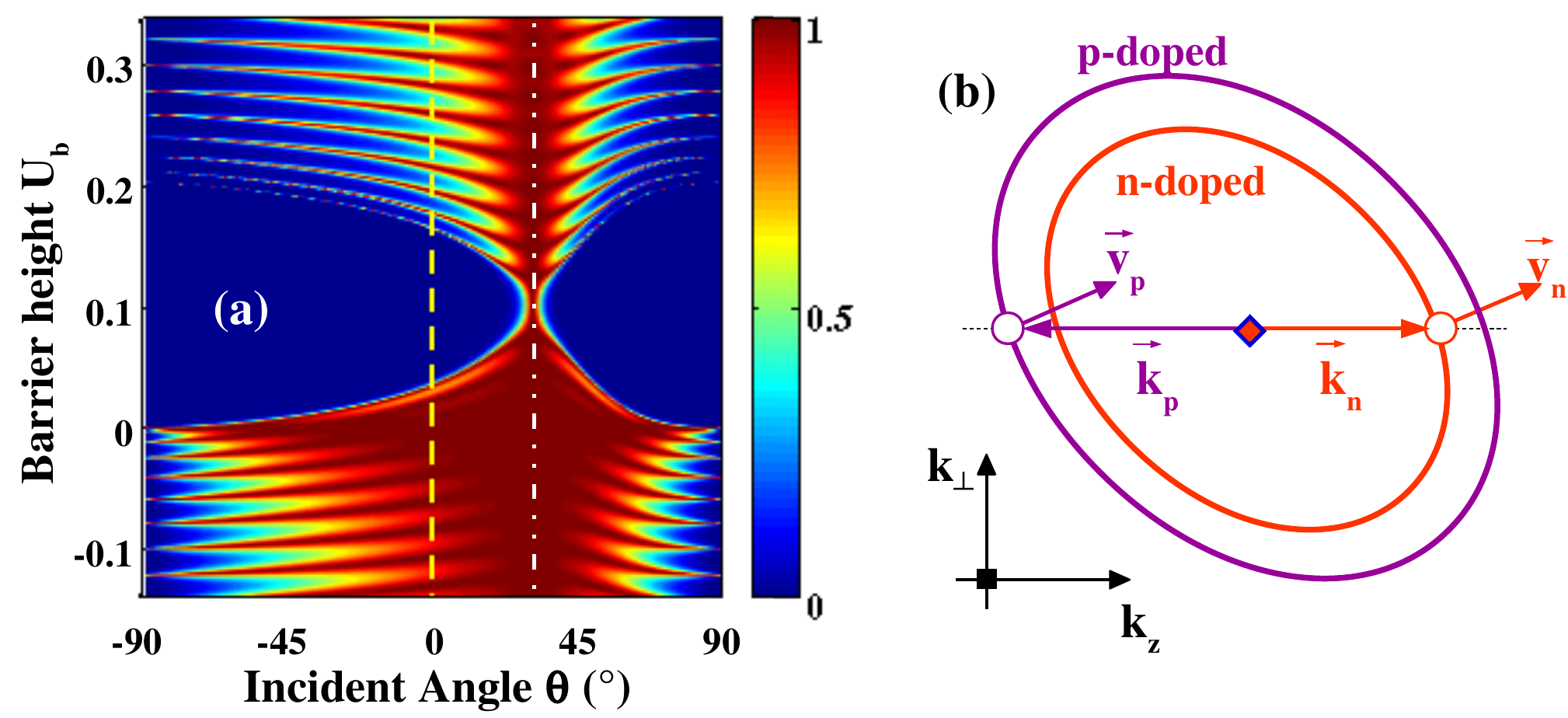}
	\caption{\textbf{The effects of anisotropic energy dispersion.} (\textbf{a}) ($U_b$-$\theta$) map of transmission probability and (\textbf{b}) bandstructure diagram illustrating the Klein tunneling process. The parameters $v_{Fx} = v_{Fz} = v_F$, $v_{Fy} = 0.5v_F$, $\gamma = 45^\circ$ (see text), $L_b = 200 a_0$, $E = 0.1$, and $\phi = 90^\circ$ are considered here.}
	\label{fig_sim9}
\end{figure}

Let us consider a simple case when a rotation $R_x (\gamma)$ of angle $\gamma$ is applied to rotate the transport direction (i.e., Oz axis) about the Ox axis. In this rotated axis, the vectors $\vec k$, ${\vec \sigma}_{\tau_c\tau_b}$ and ${\vec v}_{\tau_c\tau_b}$ are determined as  
\begin{eqnarray}
\vec k &=& k_x \vec e_x + (k_y \cos\gamma + k_z \sin\gamma) \vec e_y + (k_z \cos\gamma - k_y \sin\gamma) \vec e_z, \\
{\vec \sigma}_{\tau_c\tau_b} &=& \tau_c\tau_b \frac{\xi_{\vec k,x} \vec e_x + (\xi_{\vec k,y}\cos\gamma + \xi_{\vec k,z}\sin\gamma)\vec e_y + (\xi_{\vec k,z}\cos\gamma - \xi_{\vec k,y}\sin\gamma)\vec e_z}{\sqrt{\xi_{\vec k,x}^2 + \xi_{\vec k,y}^2 + \xi_{\vec k,z}^2}} \\
{\vec v}_{\tau_c\tau_b} &=& \tau_b \frac{\eta_{\vec k,x} \vec e_x + (\eta_{\vec k,y}\cos\gamma + \eta_{\vec k,z}\sin\gamma)\vec e_y + (\eta_{\vec k,z}\cos\gamma - \eta_{\vec k,y}\sin\gamma)\vec e_z}{\sqrt{\xi_{\vec k,x}^2 + \xi_{\vec k,y}^2 + \xi_{\vec k,z}^2}}
\end{eqnarray}
Here, ${\vec e}_{x,y,z}$ are the unit vectors in the x-, y-, and z-axes while $k_{x,y,z}$ indicate the x-, y-, and z-components of $\vec k$, respectively, in the original axis, i.e., when the rotation $R_x (\gamma)$ is not applied. Actually, the Klein tunneling has been demonstrated to achieve with the pseudo-spin conservation, that is usually obtained when $\vec k_\perp = 0$. In the rotated axis, such the condition is satisfied when $k_x = 0$ and $k_y\cos\gamma + k_z\sin\gamma = 0$. Accordingly, the pseudo-spin and group velocity of the transmitted states in both $p-$ and $n-$doped zones are given by
\begin{eqnarray}
\vec \sigma_K &=& \tau_c \vec e_y \frac{v_{Fz} - v_{Fy}}{\sqrt{v_{Fy}^2 \sin^2\gamma + v_{Fz}^2 \cos^2\gamma}}\frac{\sin2\gamma}{2} + \tau_c \vec e_z \frac{v_{Fy} \sin^2\gamma + v_{Fz} \cos^2\gamma}{\sqrt{v_{Fy}^2 \sin^2\gamma + v_{Fz}^2 \cos^2\gamma}} \\
\vec v_K &=& \vec e_y \frac{v_{Fz}^2 - v_{Fy}^2}{\sqrt{v_{Fy}^2 \sin^2\gamma + v_{Fz}^2 \cos^2\gamma}}\frac{\sin2\gamma}{2} + \vec e_z \sqrt{v_{Fy}^2 \sin^2\gamma + v_{Fz}^2 \cos^2\gamma} %\textrm{sign}(k_z \cos\gamma).
\end{eqnarray}
Thus, the pseudo-spin $\vec \sigma$ with $\vec k_\perp = 0$ is indeed conserved and a refractionless character is simultaneously achieved, which is shown below to correspond to the Klein tunneling transmission. 

We now consider the anisotropy with $v_{Fy} \neq v_{Fz}$. It is shown that if the transport direction is along one of the symmetry directions mentioned above (i.e., $\gamma = n\,90^\circ$ with $n=0,1,...,3$), this anisotropic character does not exhibit any significant effect on the transport pictures, i.e., $\vec v_K$ is perfectly parallel to the Oz axis (i.e., $\theta_K = 0^\circ$) and accordingly the transmission probability is qualitatively similar to that obtained for the isotropic dispersions \cite{hill17}. However, if the transport direction is not parallel to any symmetry direction (i.e., $\sin2\gamma \neq 0$), qualitative changes occur: the Klein tunneling is still preserved but is observed in an oblique direction determined correctly by the velocity $\vec v_K$ in the equation (S22) and accordingly a deflection of Fabry-P\'{e}rot resonances is also observed (see Fig.\ref{fig_sim9}.a), similar to that obtained under the tilting effects. This result is simply a consequence of the non-collinearity of $\vec v$ and $\vec k$ as demonstrated above and schematically illustrated in Fig.\ref{fig_sim9}.b. Here, another exceptional feature should be emphasized. In all cases (i.e., including also the tilted dispersions) without the anisotropic character, the pseudo-spin of Klein tunneling beams is always parallel to the transport direction, i.e., ${\vec \sigma}_K = \tau_c {\vec e}_z$. However, this picture changes with the effects of anisotropic energy dispersion as the pseudo-spin ${\vec \sigma}_K$ is determined by the equation (S21). Moreover, in contrast to the tilting effects, the anisotropic character does not include a momentum shift of the Fermi surfaces in the hetero-doped structures and results in the same deflection of transparent directions for the $W$ and $W'$ cones.

\end{document}